\documentclass{nature}

\usepackage{amsfonts,amssymb,amsmath,amsthm,amscd}

\usepackage{color}
\usepackage{tikz}
\usetikzlibrary{arrows,decorations.pathmorphing}
\usepackage{framed}

\newcommand{\R}{\mathbb{R}}

\newcommand{\M}{\mathcal{M}}

\newcommand{\dout}{d_{\text{out}}}
\newcommand{\din}{d_{\text{in}}}
\newcommand{\dpast}{\Omega_{\text{past}}}

\renewcommand{\sp}{\mathrm{sp}}

\newcommand{\<}{\langle}		
\renewcommand{\>}{\rangle}		

\title{The experiment paradox in physics}

\author{Micha\l\ Eckstein${}^{1,2}$, Pawe\l\ Horodecki${}^{3,4}$}

\makeatletter
\let\saved@includegraphics\includegraphics
\AtBeginDocument{\let\includegraphics\saved@includegraphics}
\renewenvironment*{figure}{\@float{figure}}{\end@float}
\makeatother

\begin{document}

\maketitle

\begin{affiliations}
 \item Institute  of  Theoretical  Physics  and  Astrophysics,
National  Quantum  Information  Centre,  Faculty  of  Mathematics,  Physics  and  Informatics,
University  of  Gda\'nsk,  Wita  Stwosza  57,  80-308  Gda\'nsk
 \item Copernicus Center for Interdisciplinary Studies, ul. Szczepa\'nska 1/5, 31-011 Krak\'ow, Poland
 \item International  Centre  for  Theory  of  Quantum  Technologies,  University  of  Gda\'nsk,  Wita  Stwosza  63,  80-308  Gda\'nsk,  Poland
 \item Faculty of Applied Physics and Mathematics, National Quantum Information Centre,
Gda\'nsk University of Technology, Gabriela Narutowicza 11/12, 80-233 Gda\'nsk, Poland
\end{affiliations}

\begin{abstract}
Modern physics is founded on two mainstays: mathematical modelling and empirical verification. These two assumptions are prerequisite for the objectivity of scientific discourse. Here we show, however, that they are  contradictory, leading to the `experiment paradox'. We reveal that any experiment performed on a physical system is --- by necessity --- invasive and thus establishes inevitable limits to the accuracy of any mathematical model. We track its manifestations in both classical and quantum physics and show how it is overcome `in practice' via the concept of environment. We argue that the scientific pragmatism ordains two methodological principles of compressibility and stability.
\end{abstract}

The methodology of physics, pioneered by Archimedes, Galileo and Newton, has been crystallising with the development of formal languages and statistical analysis. Its ``unreasonable effectiveness''\cite{Wigner} is founded on two basic assumptions: Firstly, physical systems are modelled via mathematical structures, which guarantee the universality and objectivity of the description. Secondly, the models are verifiable via a comparison of their predictions against empirical data.

The question whether there exists an `ultimate' mathematical model of physical reality (or an overarching ``law of physics''), and whether it is intelligible, is controversial and has long been the subject of philosophical debate\cite{Wheeler83,Deutsch86,Weinberg94,HawkingGodel,Heller2009}.  This problem, however, seems irrelevant for the practice of doing physics. Indeed, because of the finiteness of the available resources --- any theory and any set of empirical data must eventually be expressed as a finite combination of some intersubjective symbols --- no model can become arbitrarily accurate. Yet, we have to assume that our ignorance is the only source of the uncertainty, for otherwise we would decree ``[\ldots] that there exist aspects of the natural world that are fundamentally inaccessible to science.''\cite{Deutsch86}

The second inexorable\cite{EllisSilk} pillar of modern physics is the falsifiability of mathematical models against empirical data. Again, in practice the hypotheses can only be confirmed at some confidence level, yet we have to assume that this is solely caused by our incapabilities. In order for the experiments to be conclusive and repeatable, we have to assure that they are free --- that is the input cannot be correlated with the studied physical system, until the experiment is actually performed. The pertinence of this assumption has been recognised only recently on the occasion of the Bell tests\cite{AspectBell2015,HallBell,Bell_quasars} demonstrating the predictive supremacy of quantum mechanics over ``hidden variables'' explanations. Heuristically, it has been voiced by Stephen Hawking and George F.R. Ellis: ``[\ldots] the whole of our philosophy of science is based on the assumption that one is free to perform any experiment.''\cite[p. 189]{HawkingEllis}

\section{Local physics}

The prodigious success of physics relies on the fact that one can probe physical systems \emph{locally}, that is abstracting from the `rest of the world'. The possibility of making a cut between a system and its `compliment' has been implicitly assumed in the scientific discourse from its dawn, but has acquired a concrete shape only at the beginning of the  20th century. Voiced by Michael Faraday and sharpened by Albert Einstein the principle of locality brought about the concept of spacetime consisting of events\cite{Haag}. The formalisation of the notion of an event gave it an operational sense and thus established a rigorous link between the mathematical theory and `real world' experiments.

The Einsteinian notion of spacetime (that is a 4-dimensional smooth Lorentzian manifold\cite{Wald}) could be and has been questioned from different standpoints\cite{Earman89,QGreview}. Nevertheless, any physical theory must eventually make reference to the \emph{effective spacetime}, in which the experimental data is being gathered and shared. This comes about as follows:

Let us firstly note that intersubjective data is but a (finite) set of values of some registers. For sake of simplicity we shall assume that the registers are bits, as any more complicated universal description of the data sets can eventually be rewritten in the binary form. Now observe that any data always comes in a sequence, that is as a totally ordered set of bits. Hence, it has an inbuilt time-ordering informing that the bit $b_1$ has been input, acquired or communicated before the bit $b_2$. Furthermore, given two data sequences one needs two different labels, say $a$ and $b$, to distinguish them. The notation $\{b_1,b_2,b_3,\ldots\}$ purports that the bit $b_2$ has been obtained at a place ``$b$'' in space and a time ``2''. Thus, any bit of the data uniquely specifies an \emph{event} -- a point in space and time. 

The events are intersubjective -- because the data is so. This is not at variance with the fact that one can relabel the bits by ascribing them some explicit spacetime coordinates, such as $\{(b_1,t_1,x_1),(b_2,t_2,x_2),\ldots\}$. However, when doing so, one has to specify a `covariance law' allowing one to unambiguously translate the local data from one coordinate chart to another. Any such covariance law must hence eventually refer to an invariant object --- call it a spacetime ---, which is common to all observers. The demand of covariance is not a falsifiable physical principle, but a methodological assumption\cite{Barbour2001} --- a prerequisite for the intersubjectivity of physics.

To accommodate the local data any physical model must include a description of the effective spacetime. But the actual purpose of a theory is to provide explanations, which go beyond the directly observed phenomena. Therefore, the effective spacetime ought to contain also the \emph{potential events}. These are not directly associated with any empirical data, but they are necessary for a consistent explanation of the data. For example, the detection of a photon is an actual event, whereas the photon emission from a star is a potential event indicated by the quantum theory. Viewed from a different angle, the potential events are associated with empty registers, denoted by~$\emptyset$, for the data. This gives justice to the slogan: ``Unperformed experiments have no results.''\cite{NoResults}

However, a conceivable explanation of the actual events in terms of the past potential ones is not a sufficient criterion for a credible theory of physical reality. Indeed, the gist of physics (and, actually, of the entire science) is the ability to \emph{predict} future events. In particular, in order to conceive technology of any sort we must be able to precisely forecast the future behaviour of the designed devices. To enable predictions a theory must firstly make a distinction between the past potential events and the future ones. More precisely, it has to specify a \emph{causal structure} telling us which of the past potential events could have influenced the actual ones and which of the future events might be affected by the present ones. For instance, in Newton's theory the future abuts the past, so one cannot exclude in principle an immediate influence of a phenomenon in some remote part of the universe on our empirical data. On the other hand, in Einstein's theory the regions of potential influence are constrained by the light cone structure.

\section{The inconsonance of local experiments}\label{sec:incons}

A \emph{prediction} of a theoretical model $\M$ is a claim about some future potential events. A sharp prediction takes the form of a conditional claim: ``If $\din^p$ was input and $\M$ is valid, then $\dout^p$ ought to be registered.'' In such a case a single experiment with $\din = \din^p$, but $\dout \neq \dout^p$ would be sufficient to falsify the model $\M$. In general, one formulates --- more modest --- \emph{statistical predictions}. These are expressed as conditional probabilities $P(\dout^p \, \vert \, \M, \din^p)$ and they require multiple independent experiments with $\din = \din^p$ to validate $\M$ at a prescribed confidence level, which we decree as satisfactory. Any two competing models $\M$ and $\M'$ of a given phenomenon ought to be discernible, $\M \not\equiv \M'$, that is there must exist at least one experimental setting for which $P(\dout^p \, \vert \, \M, \din^p) \neq P(\dout^p \, \vert \, \M', \din^p)$. Let us stress that at the level of falsification it is irrelevant whether the model $\M$ is fundamentally probabilistic --- as quantum models are ---, or effectively statistical --- as a result of ignorance of some of the system's aspects, for instance its microscopic structure.

For an experiment to be trustworthy, one has to warrant that the input data is \emph{free}, that is independent of the history of the physical system at hand. Concretely, we have to ensure the statistical independence of $\din$ and past states $\dpast$ pertaining to the system. The set $\dpast = \dpast(\M)$ involves both the actual events associated with the existing data, as well as the past potential events entailed by the model $\M$. In other words, \emph{a model $\M$ must not imply correlations between} $\dpast$ \emph{and the potential future events related with} $\din$. For if it would do so, it would induce a statistical bias in what could be tested and how, hence \emph{a priori} excluding a part of physical reality from our empirical cognition. If multiple experiments with different $\din$'s are performed, so that $P(\din)$ can be defined a posteriori, we can express the demand of freedom as $P(\din \, \vert \, \dpast) = P(\din)$.

Any experiment has to involve at least one free bit of input data, as the experiment might or might not be actually performed. Note that ``experiment not done'' is indeed an objective information, which corresponds to a definite event $\din = 0$. In this case, the event associated with $\dout$ remains potential, $\dout = \emptyset$, as ``Unperformed experiments have no results.''\cite{NoResults} On the other hand, if the experiment was done, $\dout$ must take a definite value. For even if the experimental apparatus did not register any signal, this mere fact corresponds to an objective information (cf. \cite{QIandGR}). 

Let us fix a model $\M$ of a chosen physical system $F$. Let us also suppose that $F$ is embedded in an environment $E$, which is \emph{not} modelled within $\M$, but does affect the experimental outcomes. This simply signifies that $\M = \M(F;E)$ includes some noise and/or free parameters. On the physical side, it means that $F$ did and does interact with $E$, which is \emph{natural}, i.e. in principle modellable.

Suppose now that an experiment with some $\din$ and $\dout$ has been performed, to check the validity of $\M$. Since $\din$ is an (intersubjective, that is `classical') information it must be physical, hence it had to be supported (or `written') in a form of `matter' $G$. Whatever model of $G$ we would consider, it must be allowed to interact with the studied physical system $F$, because it actually did so in the experiment just performed.

Let us suppose that the model $\M$ did encompass the interaction of $F$ and $G$. If $G$ is a part of $F$, that is $\M(F,G;E) \equiv \M(F;E)$, then, clearly, $\din$ could not have been free. Suppose then that $G$ was included in the environment part: $\M(F;E,G) \equiv \M(F;E)$. But then we admit that the entire `experiment' was actually a natural, that is modellable, phenomenon. Consequently, even if no correlations between $\din$ and $\dpast$ were assumed within $\M$, there exists an extended model $\overline{\M} = \overline{\M}(F,E)$ providing a natural explanation of the entire `experiment' hence, a statistical dependence between $\din$ and $\overline{\Omega}_{\text{past}} = \overline{\Omega}_{\text{past}}(\overline{\M})$.

Therefore, in order to guarantee the freedom of $\din$, we have to assume its independence in \emph{any} conceivable model
\begin{align*}
P(\din \, \vert \, \dpast(\M)) = P(\din), \quad \text{ for all } \M. 
\end{align*}
In other words, the experimental input $\din$ has to be \emph{random} (cf. \cite{RandomnessAmp2012}). But then, once the experiment was performed, we must update the model to take into account the interaction of $F$ with $G$, symbolically: $\M'(F,G;E) \not\equiv \M(F;E)$. Note that such a change has \emph{global} consequences --- it affects, in general, not only the future potential events, but also the past ones, because $\dpast = \dpast (\M)$. 

This is what we might call the \emph{experiment paradox}: We must assume that the experimental input is free in order to perform credible tests of theoretical models, but then we allow for `non-physical' interventions,  which are --- by assumption --- not modellable. 

\section{Faces of the experiment paradox}

We now unravel the manifestations of the experiment paradox in well-established physical theories.

A convenient universal framework for both classical and quantum mechanics uses the language of \emph{states} -- encoding the properties of a given physical system $F$ and \emph{observables} -- measurable physical quantities\cite{Strocchi}. Any observable $A$ has a spectrum $\sp(A) \subset \R$, that is a set of possible measurement outcomes and any state $\rho$ defines a probability distribution $\mu_{\rho,A}$ over the set $\sp(A)$.

Models of local physical phenomena are formulated in terms of dynamical equations
\begin{align}\label{evo}
f(\rho(t),t) = 0, \quad \text{for} \quad t \in [0,T],
\end{align}
where $t$ is a time parameter and $f$ is functional (typically, a linear differential operator) acting on the space of states. Such a model specifies the time-evolution of the system's state $\rho(t)$ from an initial condition
\begin{align}\label{const}
g(\rho(t),t) \vert_{t=0} = 0,
\end{align}
determined by a (collection of) functionals $g$. It involves, in particular, the initial state $\rho(0)$.

The predictions of the model \eqref{evo} are then formulated as follows: If the system was initially described by \eqref{const} and an observable $A$ was measured at a time $t > 0$, then an outcome $a\in\sp(A)$ will be obtained with probability $\mu_{\rho(t),A}(a)$\footnote{If the observable $A$ has a continuous spectrum, then the `outcome' is specified within some interval $[a-\delta,a+\delta]$ and the probability is given by $\mu_{\rho(t),A}([a-\delta,a+\delta])$.}. Hence, to test a model determined by equation \eqref{evo} one has to \emph{prepare} the system in an initial condition \eqref{const} and then measure an observable $A$ at some time $t \in (0,T]$. Multiple experiments with different inputs would tell us whether the predicted probabilities match the observed ones.

Note that the experimental input listed above is indeed free within model \eqref{evo}, because the latter specifies neither the initial conditions (The model does specify the admissible forms of the initial conditions, but not the numerical values.) nor the observable $A$ and the measurement time $t$. However, we have to admit that the studied system had been in \emph{some} state (for instance, the vacuum state) before it was prepared by the experimentalist.

Quantum theory, as opposed to classical mechanics, provides a formal operation corresponding to the reset of system's state -- the von Neumann projective measurement. But, the admission of projective measurements leads to the notorious measurement paradox (see Box 1).

To circumvent the `resetting problem' we could construct an extended model
\begin{align}\label{evo2}
f(\rho(t),t) = 0, \quad \text{for} \quad t \in [t_0,T],
\end{align}
describing the evolution of the system from some earlier time $t_0 < 0$ until $T$. Then, we assume that its dynamics has been \emph{perturbed} 
\begin{align}\label{evoj}
f(\rho(t),t) = j(t), \quad \text{for} \quad t \in [t_0,0),
\end{align}
with a suitable \emph{source} $j$, so that the desired condition \eqref{const} at $t=0$ is met, regardless of the primordial system's initial conditions at $t=t_0$.

But, clearly, models \eqref{evo2} and \eqref{evoj} are different and the introduction of a source term is invasive. Had the experiment not been performed, the object would evolve according to equation \eqref{evo2} rather than \eqref{evoj}. If, on the other hand, we attempt to model the source itself we lose (or rather, shift to another level of complexity) its tunability --- hence the `preparation paradox' (see Box 1).

The same line of reasoning could be followed in the Heisenberg picture, in which the system's state remains steady, but the observables evolve in time. Let us also note that the source term $j$ need not be a function --- it could be, for instance, a time-dependent Hamiltonian appended to the Schr\"odinger equation of a given system. 

In conclusion, regardless of whether the theory entails that the measurement --- i.e. information acquisition --- disturbs the system or not, the preparation procedure is always invasive.

Fortunately, the experimental outcomes typically depend very weakly on how the system has been prepared. The `triggering effects', that is the details of the source $j$, can usually be alleviated below the noise level shaped by the uncontrolled interaction of the system with its environment.

\begin{figure}[h]
\begin{center}
\includegraphics[scale=0.35]{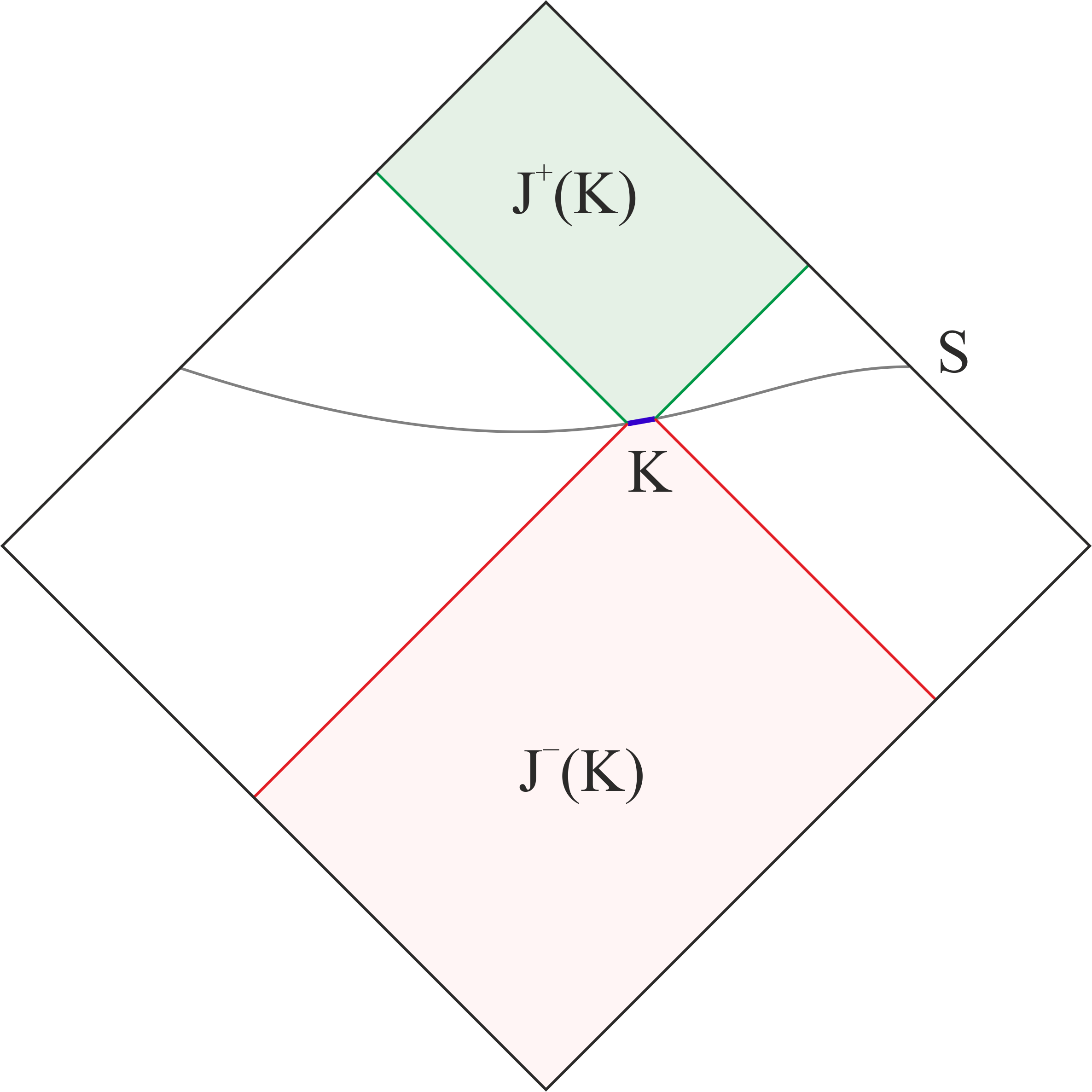}
\end{center}
\caption{\label{fig:cosmos}The conformal diagram for the Minkowski spacetime. The field content of the entire spacetime is uniquely determined by initial data imposed on a Cauchy hypersurface $S$. Consequently, a free intervention effectuated in the region $K$ induces a change both in the causal future $J^+(K)$ and the causal past $J^-(K)$ of $K$. More generally, the outer diamond could serve as an illustration for the maximal Cauchy development of the hypersurface $S$.}
\end{figure}

The experiment paradox is, however, more salient in the cosmological context, which does not leave room for any environment. Modern cosmological models are formulated in the framework of field theory. Let us emphasise that the fields do not evolve per se --- a solution to field equations specifies the field content in the entire (effective) spacetime. Therefore, any disturbance coming `from outside' would effectuate a global change. In other words, a local terrestrial experiment affects both future and past states of the Universe (see Figure \ref{fig:cosmos}). Note also that cosmological observations are indeed genuine experiments for, firstly, they might but need not be effectuated and, secondly, they involve a number of tunable free parameters, such as the telescope's location and direction or electromagnetic spectrum sensitivity range.

As an illustration, let us consider a cosmological model based on Einstein's equations
\begin{align}\label{Einstein}
G_{\mu\nu} = \tfrac{8 \pi G}{c^4} T_{\mu\nu},
\end{align}
with a matter energy--momentum tensor $T_{\mu\nu}$ (possibly including the ``dark energy'', i.e. the cosmological constant term $\Lambda g_{\mu\nu}$). The geometrical Bianchi identity $\nabla^{\mu} G_{\mu\nu} = 0$ implies the local covariant conservation of energy and momentum $\nabla^{\mu}  T_{\mu\nu} = 0$\cite{Wald}. But, if an `external' source term $j_{\nu}$ is introduced into \eqref{Einstein}, the conservation law is violated, $\nabla^{\mu}  T_{\mu\nu} = -j_{\nu}$, explicitly breaking general covariance. In other words, if one introduces into the universe some information which was not there, one creates \emph{ex nihilo} a local source of energy--momentum. 

In quantum field theory, whereas the energy and momentum need not be conserved locally, the suitable expectation values ought to be conserved. Concretely, if $\hat{T}_{\mu\nu}$ is the energy--momentum operator constructed from quantum matter fields, then 
\begin{align}\label{cons}
\nabla_{\mu} \< \psi \vert \hat{T}^{\mu\nu} \vert \psi \> = 0
\end{align}
should hold\cite{Bertlmann} for any state vector $\vert \psi \>$. The introduction of a, possibly quantum, source $\hat{j}$ violates the constraint \eqref{cons} leading to the Einstein anomaly and, eventually, to the breakdown of general covariance\cite{Bertlmann} (see also \cite{Bednorz}).

In order to perceive the experiment paradox from the perspective of `cosmic evolution' we firstly need to choose a time function --- that is an observer ---, which fixes an effective splitting of the global spacetime into space and time\cite{Wald}. Secondly, one has to assure that equations \eqref{Einstein} allow for a well-defined Cauchy problem\cite{HR09}. The latter consists in imposing initial data on a time-slice, say at observer's time $t=0$, and studying its (maximal) hyperbolic development (see Figure~\ref{fig:cosmos}). This guarantees that both past and future field configurations are uniquely derived from the imposed initial data. The objectivity of the evolution is guaranteed by general covariance, which enables unequivocal transcription of the time-slice field configurations for different observers.

Now, a free perturbation, or an abrupt change of initial data on a time-slice in a region $K$ of space inflicts a change in both causal future $J^+(K)$ and causal past $J^-(K)$ of $K$. The problem persists in the context of quantum field theory, because of the ``time-slice axiom''\cite{Haag}. This is independent from the fact that projective measurements are as harmful to quantum field theory as they are to the non-relativistic quantum theory.

\section{Consequences for foundations of physics}

If we define the `fundamental level of physics' as being \emph{in principle} subject to both arbitrarily accurate modelling and experimental probing, then it does not exist --- because of the unveiled paradox. Furthermore, the assumption about the existence of `ontic' random events is a methodological necessity (see Box 2). Yet, there is no decisive procedure to check \emph{post factum} whether an event was random or not.

The experiment paradox has profound consequences for the philosophy of science, which shall be discussed elsewhere\cite{EHH_phil}. Nevertheless, from the practical point of view, one may adopt the perspective that ``all models are wrong, but some of them are useful.''\cite{Box}

The `usefulness' of theoretical models is quantified by their explanatory and predictive power. These rely on two key properties: \textit{compressibility} and \textit{stability}. The former means that we can describe large sets of empirical data within a tight theoretical scheme based on several overarching rules. Notwithstanding, the theory itself might need to be expressed in terms of sophisticated mathematical structures\cite{Dirac63}. By ``stability'' we understand the independence of the laws of nature from the testing procedures. It guarantees the repeatability of experiments and, eventually, enables the construction of trustworthy devices.

As expressed in a compressed quote from Albert Einstein: ``Everything should be made as simple as possible, but no simpler.''\cite{EinsteinQuote}

\begin{addendum}
 \item We are grateful to Michael Heller, Ryszard Horodecki and Tomasz Miller for inspiring discussions and enlightening comments on the manuscript.
 
 The work of ME was supported by the National Science Centre in Poland under the research grant Sonatina (2017/24/C/ST2/00322). PH acknowledges support by the Foundation for Polish Science through IRAP project co-financed by EU within Smart Growth Operational Programme (contract no. 2018/MAB/5).
\end{addendum}

\pagebreak

\begin{framed}
\noindent \textbf{BOX 1: Chains of causal reasoning}

1) The measurement paradox

\vspace*{-0.2cm}
\begin{center}
\includegraphics[scale=0.65]{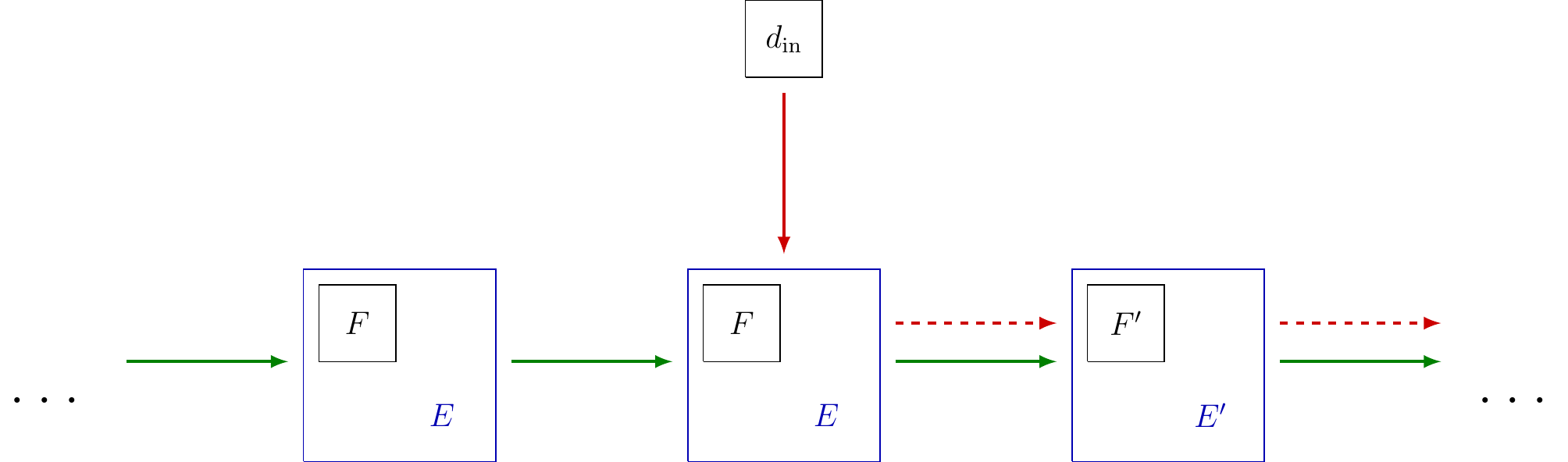}
\end{center}

Let $F$ be a quantum system described by the quantum state $\rho$ and suppose that we choose to measure an observable $A$, according to our free input $\din$. Then, the standard von Neumann postulate of quantum mechanics implies that after the measurement the system's state jumps abruptly to $\rho'$ -- one of the (pure) eigenstates of the observable $A$. Such a `non-physical' intervention can be given a natural explanation by embedding $F$ in a (quantum) environment $E$ and invoking the formal equivalence of projective measurements on $F$ with a unitary evolution on $F \otimes E$\cite{vonNeumann}. But then we face the notorious Wigner's friend paradox\cite{WignerFriend} and are eventually forced to conclude\cite{RennerMW} that no definite $\dout$ can ever be consistently produced. 

2) The preparation paradox

\vspace*{-0.2cm}
\begin{center}
\includegraphics[scale=0.65]{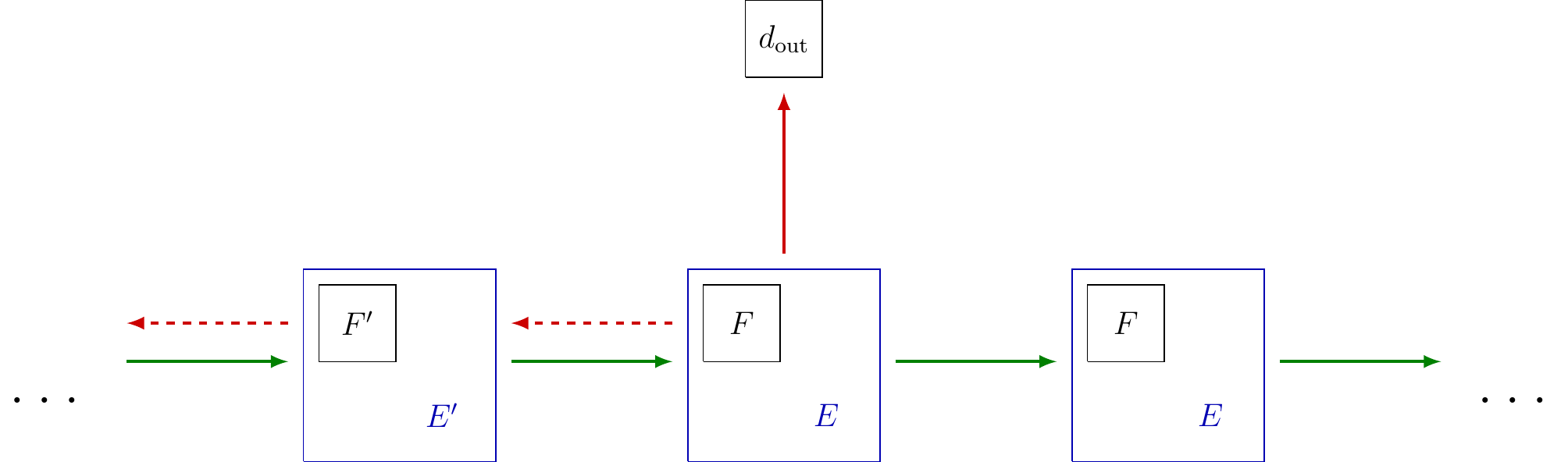}
\end{center}

The preparation paradox is the `mirror' version of the measurement paradox. Suppose that we have obtained an output $\dout$ from an experiment performed on the system $F$. By seeking a `purely natural' explanation of $\dout$ we have to embed $F$ in an environment $E$, the interaction with which caused $\dout$. But then no $\din$ has ever occurred and we have never actually prepared the system in any way.

3) Superdeterminism

\begin{center}
\includegraphics[scale=0.65]{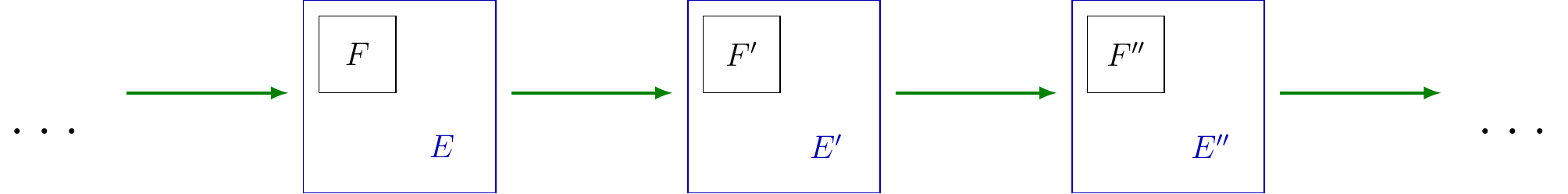}
\end{center}

One could maintain that we never actually prepare physical systems or perturb them -- we just observe them evolving without effectuating any disturbance. But such a superdeterministic viewpoint excludes \emph{a priori} the possibility of any interaction with the `physical world', in particular it disallows any experiments. This firstly annihilates the explanatory power of science and, secondly, it is highly unpractical for it excludes \emph{a priori} the existence of devices functioning according to our inputs.

4) Scientific Pragmatism

\vspace*{-0.2cm}
\begin{center}
\includegraphics[scale=0.65]{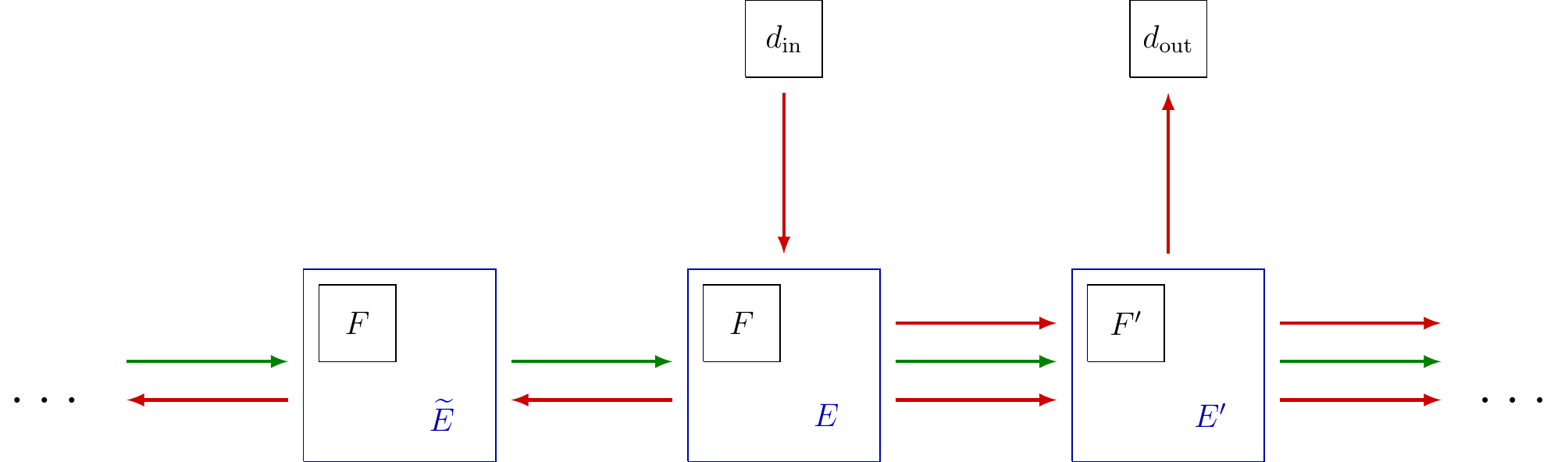}
\end{center}

In scientific practice we assume that our interactions with the studied system $F$ encoded in $\din$ do not have natural, i.e. modellable, causes and that the obtained information $\dout$ is always definite and objective. In order to save the model's consistency we have to warrant that our interventions do not affect the past states of the system $F$. To this end we need to embed it in a suitable environment $E$, which absorbs the `retrocausal' effects ($E \to \widetilde{E}$) and enables a consistent description of system's history by multiple observers. Whether we wish our interventions to affect the system's future states ($F \neq F'$, $E = E'$) or not ($F = F'$, $E \neq E'$) depends on whether we work in the observational paradigm (as, for instance, in cosmology) or in the engineering one.

\end{framed}

\pagebreak

\begin{framed}
\noindent \textbf{BOX 2: Random events in nature}

Any experiment requires at least one free, i.e. \emph{random} bit (see Section \ref{sec:incons}). Quantum theory implies that some of the events related to projective measurements are indeed random, hence quantum phenomena could be used as `sources of randomness'. However, in order to test the quantum theory itself we need to perform an experiment, the free input of which does not rely on quantum theory. In general, we are always bound to \emph{assume} the freedom of some events in order to trust the experiments, as there exist no decisive procedure to check whether a finite binary sequence comes from a `source of randomness' or a complex deterministic algorithm\cite{Claude}.

This insight uncovers a captivating similarity between theories and experiments. Any formal theory is based on a collection of axioms, from which theorems are deduced. The axioms cannot be ultimately proven true or false --- this is a general fact stemming from the limiting theorems in formal systems. In practice, we assume the axioms to be true and check whether such a premise is useful for deriving new results. Similarly, any experiment is based on a set of free bits (the `input'), which facilitate an explanation --- within a theoretical scheme --- for the `output'. We do not ask whether these bits were `truly random' or not, but rather if the assumption about their freedom is useful or not.

Let us illustrate this observation with two concrete examples:

Consider an experiment consisting of multiple tosses of a coin. The assumption that its binary outcomes are random is fairly useful\cite{Oversimplifying}. Yet, a coin is admittedly a macroscopic object following a definite trajectory determined by the gravitational attraction and air drag. Hence, we \emph{could} in principle establish a model of the coin toss, which would `explain' the outcomes. Clearly, such a model would be very complex and would rely on a number of unknown parameters. Furthermore, it would likely be unstable -- new `experiments' (or, rather, `observations') would require ad hoc adjustments. In consequence, although we could provide a deterministic model of a coin toss, it would be fairly useless.

Let us now turn to the Bell test\cite{BellThm,AspectBell2015} -- a foundational experiment aimed at demonstrating the intrinsic randomness of quantum measurements. In a typical scenario two parties independently perform measurements on a shared pair of entangled particles. Assuming that the ``locality'' and ``fair sampling'' loopholes have been closed\cite{AspectBell2015}, the Bell--CHSH theorem\cite{BellThm,CHSH} says that if the measurement outcomes are determined by a ``hidden variable'' $\lambda$ \emph{and} the settings of the parties' devices are free, that is uncorrelated with $\lambda$, then a certain measure of correlations $S$ between the outcomes is bounded, $S \leq 2$. Yet, numerous experiments have shown with a high statistical significance that the value of $S$ exceeds 2, reaching the quantum bound\cite{Cirelson} $S = 2\sqrt{2}$. Consequently, one could conclude that there is no ``hidden variable'' explanation and the outcomes are `truly random', as implied by quantum mechanics. Alternatively, one could give up the ``measurement independence'' assumption and provide a fully deterministic explanation\cite{HallBell}. However, the adequacy of such a `hidden variable' model is highly questionable, as, for instance, it would require the triggering of a common cause mechanism at the early stages of the Universe's evolution\cite{Bell_quasars}.

\end{framed}

\bibliography{causality_bib.bib}




\end{document}